\documentclass[10pt,final,journal]{IEEEtran}

\pdfoutput=1

\usepackage{graphicx}
\usepackage{siunitx}
\usepackage{subcaption}

\begin{document}
\title{Characterization of Novel Thin N-in-P Planar Pixel Modules for the ATLAS Inner Tracker Upgrade}
%
%

\author{\IEEEauthorblockN{Julien-Christopher Beyer, Alessandro La Rosa, Anna Macchiolo, Richard Nisius and Natascha Savic} \\ \IEEEauthorblockA{Max Planck f{\"u}r Physik (Werner-Heisenberg-Institut) \\ F{\"o}hringer Ring 6, DE-80805 M{\"u}nchen, Germany \\ \vspace{2mm} Email: jbeyer@mpp.mpg.de}
\thanks{Manuscript received November 29, 2016.}
}

\maketitle
\pagestyle{empty}
\thispagestyle{empty}

\begin{abstract}
The ATLAS experiment will undergo a major upgrade of the tracker system in view of the high luminosity phase of the LHC (HL-LHC) to start operation in 2026. The most severe challenges are to be faced by the innermost layers of the pixel detector which will have to withstand a radiation fluence of up to 1.4x\boldmath{$10^{16}\,$}n$_\text{eq}$/cm\boldmath{$^{2}$}. Thin planar pixel modules are promising candidates to instrument these layers, thanks to the small material budget and their high charge collection efficiency after irradiation. Sensors of 100-\SI{200}{\micro \meter} thickness, interconnected to FE-I4 read-out chips, are characterized with radioactive sources as well as testbeams at the CERN-SPS and DESY. The performance of sensors irradiated up to a fluence of \boldmath$5\times 10^{15}\,$n$_\text{eq}$/cm$^{2}$ is compared in terms of charge collection and hit efficiency. Highly segmented sensors are a challenge for the tracking in the forward region of the pixel system at the HL-LHC. To reproduce the performance of 50x\SI{50}{\square \micro \meter} pixels at high pseudo-rapidities, FE-I4 compatible planar pixel sensors are studied before and after irradiation in beam tests at high incidence angle (\SI{80}{\degree}) with respect to the short pixel direction. Results on cluster shape and hit efficiency will be shown.
\end{abstract}

\begin{IEEEkeywords}
ATLAS, HL-LHC, radiation-tolerant detectors, silicon pixel detectors
\end{IEEEkeywords}




\section{Introduction}
\IEEEPARstart{T}{he} latest module technology used in the Inner Detector (ID) of the ATLAS experiment \cite{atlasexperiment} is employed in the Insertable B-Layer (IBL) Phase-0 upgrade \cite{IBL_paper} of the Pixel Detector \cite{pixeldetectorref}. The planar sensors covering the central part of the IBL have a thickness of \SI{200}{\micro \meter}, employ n$^+$-in-n technology and have a pixel cell size of $50\times$\SI{250}{\square \micro \meter}.
\\
The Phase-0 upgrade is the first of a series of upgrades to the ID. The final stage will be the complete exchange of the ID with a new detector called Inner Tracker (ITk). This upgrade will be necessary as the HL-LHC will operate at five times the luminosity of the LHC. The accompanying higher track density will require smaller pixel cells to keep the occupancy at the present level. For this purpose a new read-out chip called RD53 chip \cite{RD53paper} is being developed by the RD53 collaboration \cite{RD53collaboration} in a \SI{65}{\nano \meter} CMOS technology with a cell size of $50\times$\SI{50}{\square \micro \meter}. The ITk concept foresees an all silicon detector composed of 5 pixel detector and 4 strip detector layers. The expected integrated luminosity of $\mathcal{L}=$ \SI{3000}{\per \femto \barn} after ten years of data taking will result in a maximum expected fluence around $1.4\times10^{16}\;$n$_\text{eq}/$cm$^{2}$ for the innermost layer \cite{radiationexpectation}, taking into account the replacement of this layer at around half of the HL-LHC data taking period.
\\
Thin n-in-p planar sensors are a promising candidate for instrumenting all pixel layers and their qualification as well as certain research aspects will be shown in the following. Section \ref{section_expmethods} will serve as an introduction to the utilized experimental methods and in section \ref{section_newmodule} new sensor designs will be discussed. Finally, section \ref{section_results} will summarize the results.


\section{Experimental methods}
\label{section_expmethods}
Various experimental methods are utilized to measure hit efficiency and charge collection efficiency. Furthermore, basic properties like breakdown and full depletion voltage are obtained from electrical characterizations, namely I-V and C-V measurements.

\subsection{Laboratory measurements}
All laboratory measurements are performed inside a climate chamber with an ambient temperature of \SI{-50}{\celsius} for irradiated modules and \SI{20}{\celsius} for not-irradiated modules. Charge-collection profiles are produced with a $^{90}$Sr $\mathrm{\beta}$-electron source placed above the module with an external scintillator trigger underneath. Additional measurements are carried out with $^{241}$Am and $^{109}$Cd sources, where characteristic $\mathrm{\gamma}$-photons are used to calibrate the measured Time over Threshold (ToT) information of the read-out chip, to the collected charge. The USBPix system \cite{USBPix} is used for data acquisition.

\subsection{Testbeam}
To determine the hit efficiency of a pixel module, testbeam measurements are performed at the testbeam facilities of DESY-II and CERN SpS. DESY-II provides a \SI{5}{\giga e \volt} electron beam and CERN SpS a \SI{120}{\giga e \volt} pion beam. In both cases, EUDET type beam telescopes \cite{eudettelescope} are used to perform single event reconstruction allowing for a precise evaluation of the point of incident on the Device Under Test (DUT). The DUT is situated between three upstream and three downstream telescope planes. The telescope reaches a point resolution of \SI{3.5}{\micro \meter} for the lower energetic electrons and \SI{2}{\micro \meter} for the higher energetic pions \cite{eudettelescope}. Measurements of irradiated modules are carried out at \SI{-40}{\celsius} whereas not-irradiated modules are operated at a temperature of \SI{15}{\celsius}, all inside a cooling box, ensuring stable temperatures. Data acquisition is done with the USBPix and RCE read-out systems \cite{RCE}. An offline reconstruction of the acquired data is executed using the EUTelescope software framework \cite{eutelescope}. Analysis of the reconstructed data is subsequently done with TBmon2 \cite{tbmon2}. The results are overall hit efficiency, in-pixel efficiency and the efficiency as a function of the distance to the sensor edge. The absolute systematic uncertainty of the calculated efficiencies was estimated in \cite{testbeamresultsjens} to be 0.3\%.


\section{Sensor layout}
\label{section_newmodule}
The sensor concept presented in this paper is composed of different technologies which are tested during the development of a new pixel detector generation. The sensor is produced with an n-in-p technology, allowing for a potential cost reduction by using a single sided process. A sensor thickness of 100-\SI{200}{\micro \meter} is foreseen. The innermost layers could be equipped with the more radiation tolerant but more fragile \SI{100}{\micro \meter} thick sensors while keeping a thickness of 150-\SI{200}{\micro \meter} for the less irradiated outer layers. Pixel cell sizes of $50\times$\SI{50}{\square \micro \meter} and $25\times$\SI{100}{\square \micro \meter} are under investigation both being compatible with the currently developed RD53 chip. Due to the fact that the new chip will only be available in 2017, all presented studies are carried out with the current generation read-out chip which has a pixel cell size of $50\times$\SI{250}{\square \micro \meter}. Different approaches will be presented to still test the features of smaller pixels.
\\
Modified punch-through structures are an important element of the new module concept, to reduce the related loss of efficiency after irradiation. In these studies, sensor productions from CiS\footnote{CiS Forschungsinstitut f{\"u}r Mikrosensorik GmbH} (Germany), MPG-HLL\footnote{Semiconductor lab of the Max-Planck society} (Germany) and ADVACAM (Finland) are investigated.

\subsection{Thin sensors}
The charge collection and hit efficiency of sensors with thicknesses ranging from \SI{75}{\micro \meter} to \SI{285}{\micro \meter} are investigated.
\begin{figure}[!htp]
    \centering
    \begin{subfigure}[b]{0.45\textwidth}
           \centering
           \includegraphics[width=\textwidth]{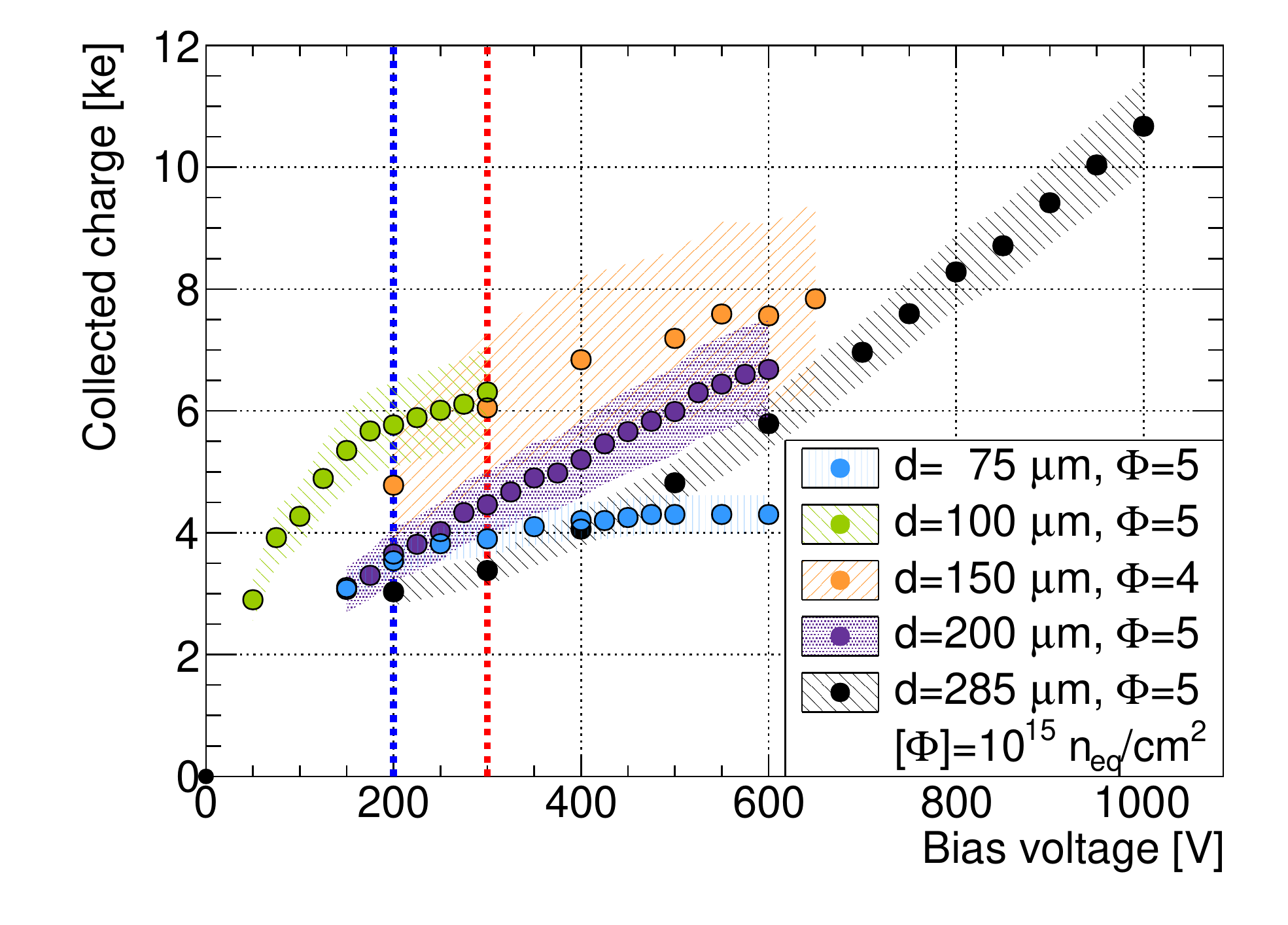}
            \caption{}
            \label{fig:1a}
    \end{subfigure}
    \begin{subfigure}[b]{0.45\textwidth}
            \centering
            \includegraphics[width=\textwidth]{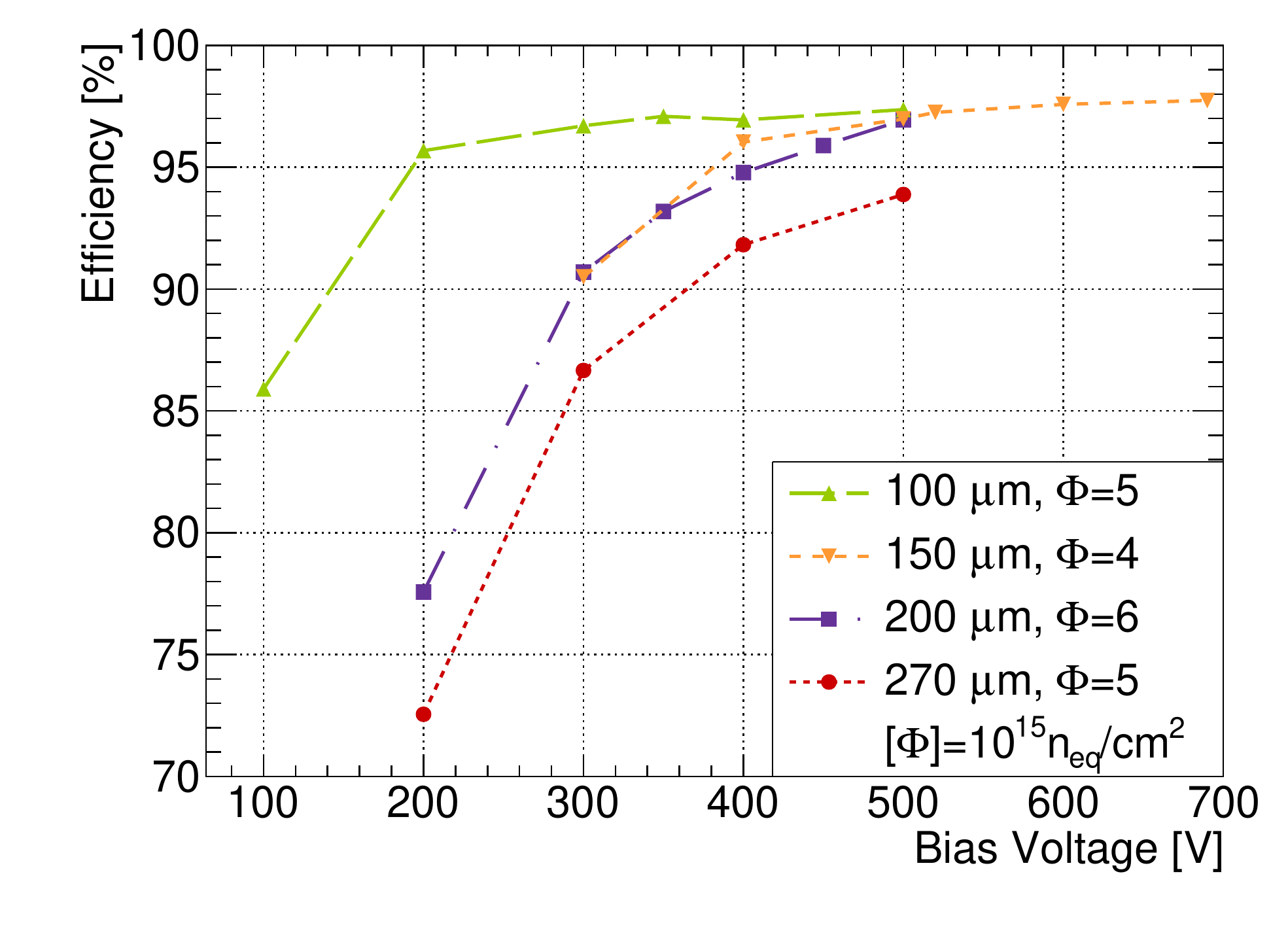}
            \caption{}
            \label{fig:1b}
    \end{subfigure}
    \caption{Characterization of irradiated sensors with thicknesses ranging from \SI{75}{\micro \meter} to \SI{285}{\micro \meter}. Figure (a) shows the collected charge as a function of bias voltage. Figure (b) shows the hit efficiency as a function of bias voltage. Both adapted from \cite{stefanothesis}.}
    \end{figure}
Figure \ref{fig:1a} shows the collected charge as a function of bias voltage for irradiated sensors. Sensors of \SI{100}{\micro \meter} and \SI{150}{\micro \meter} show the highest charge collection at moderate voltages of 200-\SI{400}{\volt}. Low bias voltages are favorable as they result in low power dissipation which allows for a lighter cooling system. Figure \ref{fig:1b} shows the corresponding hit efficiency as a function of bias voltage. Sensors of \SI{100}{\micro \meter} and \SI{150}{\micro \meter} again show the best performance at moderate voltages.

\subsection{Optimized sensor layout}
\begin{figure}[!htp]
\centering
\includegraphics[width=8cm]{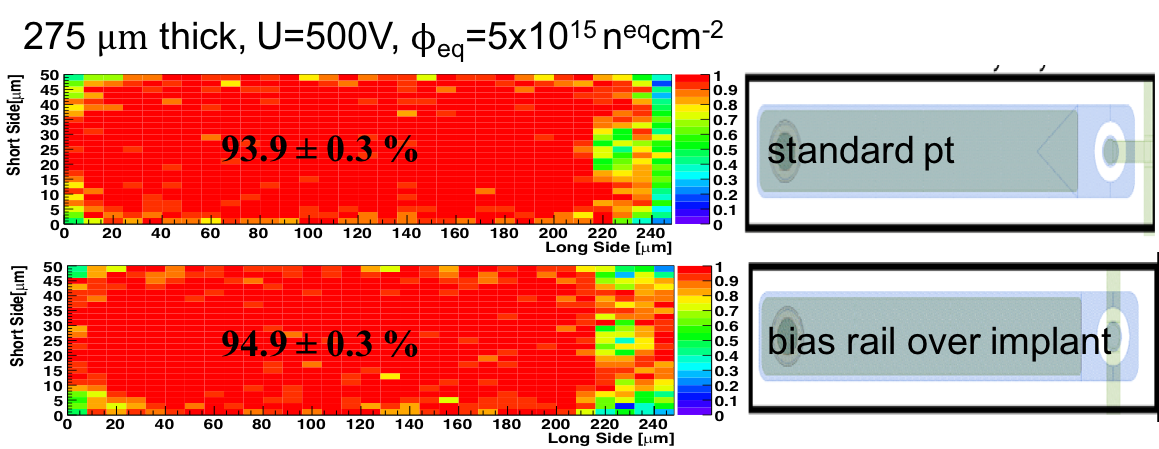}
\caption{In-pixel hit efficiency of an irradiated ($\Phi_\text{eq}=5\times10^{15}\;$n$_\text{eq}$/cm$^{2}$) \SI{275}{\micro \meter} thick module, biased at \SI{500}{\volt}. The top figure shows the efficiency for a standard punch-through structure, whereas the lower figure shows the efficiency for a modified punch-through structure with the bias rail being guided over the pixel implant instead of in between two pixel columns.}
\label{fig:2}
\end{figure}

\begin{figure}[!htp]
\centering
\includegraphics[width=6cm]{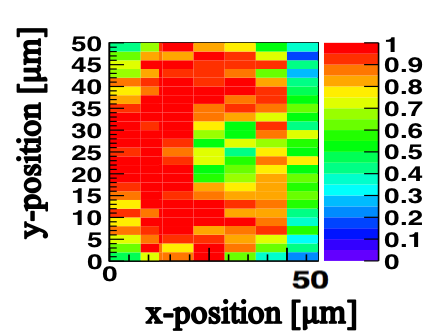}
\caption{In-pixel hit efficiency cut-out for a hypothetic $50\times$\SI{50}{\square \micro \meter} pixel cell. The cut-out is obtained by combining the left \SI{10}{\micro \meter} and the right \SI{40}{\micro \meter} of the standard punch-through hit efficiency map of Fig. \ref{fig:2}.}
\label{fig:3}
\end{figure}

\begin{figure}[!htp]
    \centering
    \begin{subfigure}[b]{0.227\textwidth}
           \centering
           \includegraphics[width=\textwidth]{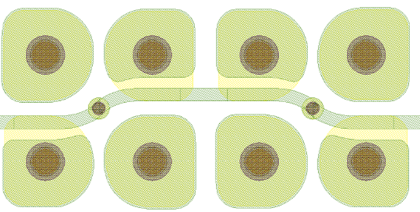}
            \caption{}
            \label{fig:32a}
    \end{subfigure}
    \begin{subfigure}[b]{0.24\textwidth}
            \centering
            \includegraphics[width=\textwidth]{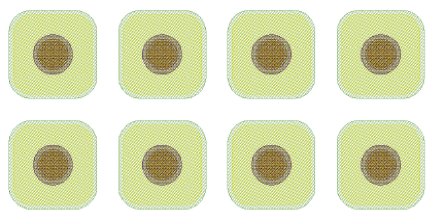}
            \caption{}
            \label{fig:32b}
    \end{subfigure}
    \caption{New layout for pixels with a cell size of $50\times$\SI{50}{\square \micro \meter}. Picture (a) shows an optimized punch-through structure with a common punch-through dot shared by four pixels and a bias rail situated as much as possible over the pixel implants. Also punch-through free designs are under investigation (b).}
         \label{fig:32}
    \end{figure}

\begin{figure}[!htp]
\centering
\includegraphics[width=8cm]{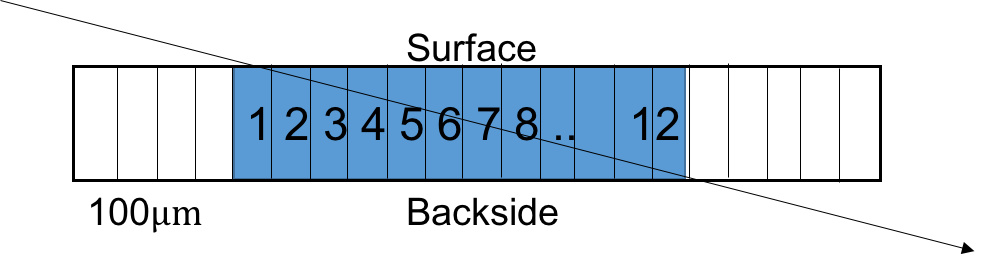}
\caption{Measurement principle of high-$\eta$ testbeam measurements. The module is tilted with respect to the beam such that it is no longer perpendicular but nearly parallel to the beam. Here, the particles enter the sensor on the side of the pixel implants (surface) and exit the sensor on the backside. The expected cluster size is given by the geometry, namely the thickness of the sensor and the incident angle. A plot showing the expected mean cluster width for different sensor thicknesses and incident angles is given in Fig. \ref{fig:42}.}
\label{fig:4}
\end{figure}

\begin{figure}[!htp]
\centering
\includegraphics[width=8cm]{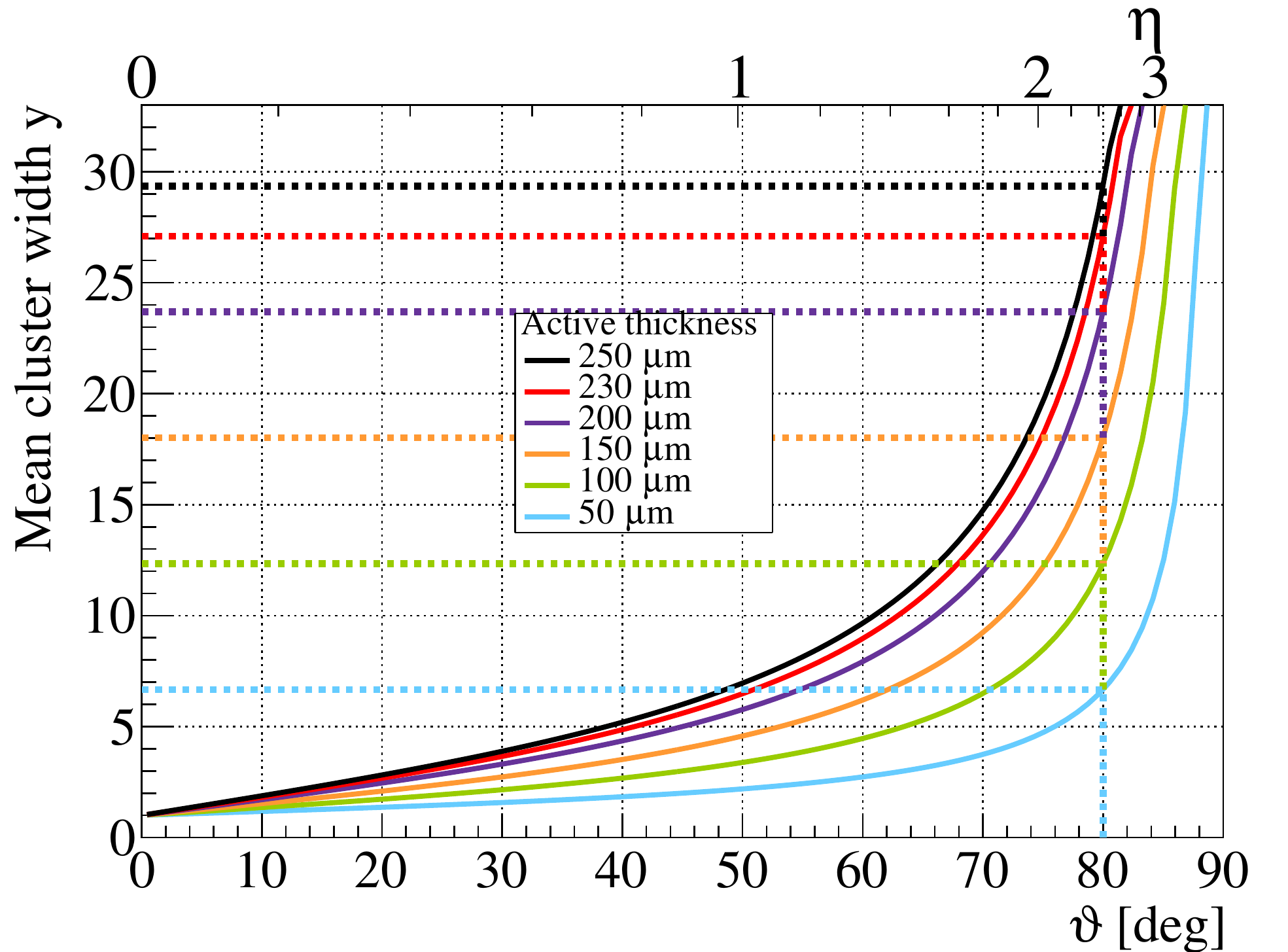}
\caption{Expected mean cluster width for different sensor thicknesses and incident angles.}
\label{fig:42}
\end{figure}

\begin{figure}[!htp]
\centering
\includegraphics[width=8cm]{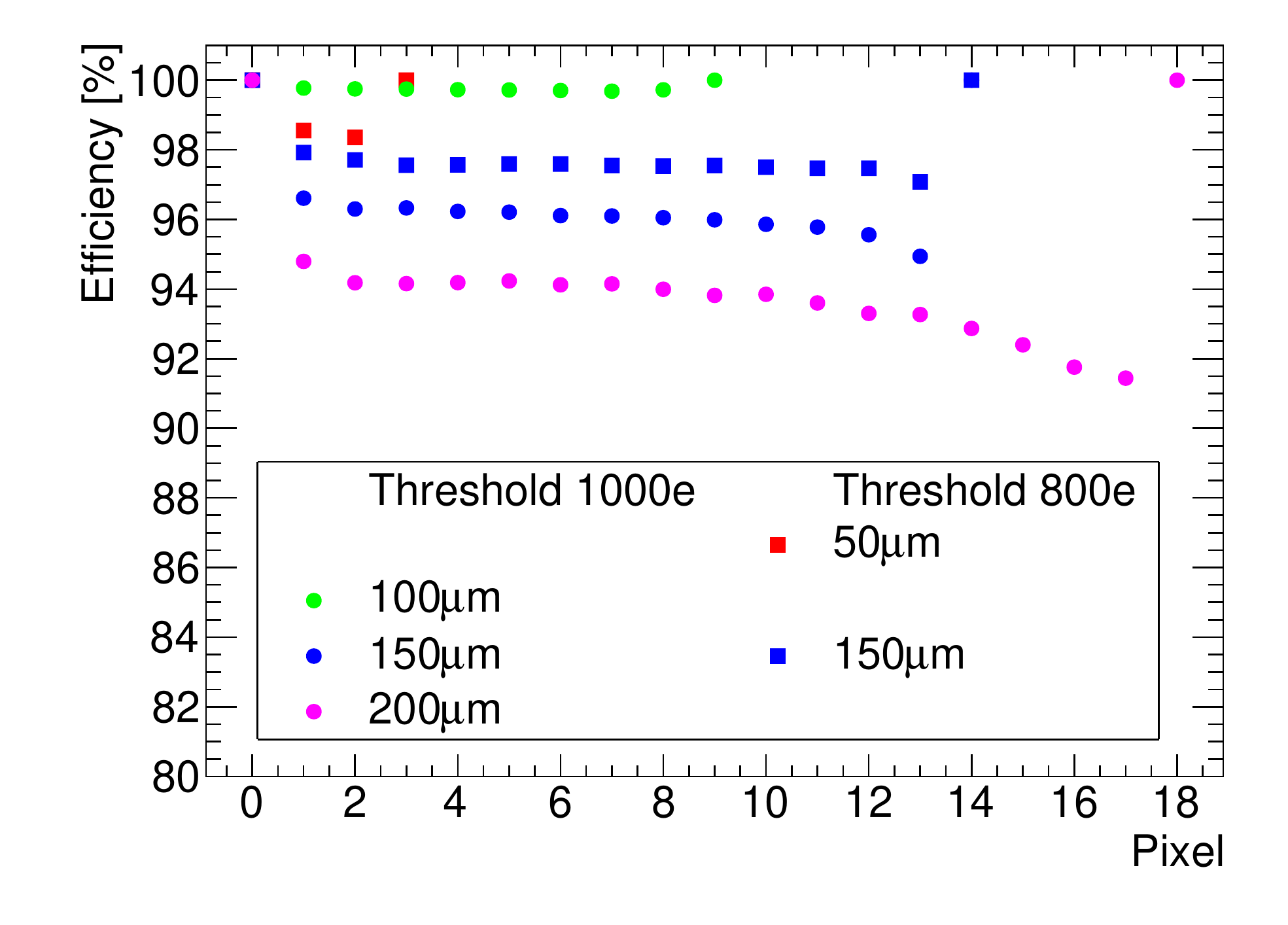}
\caption{Efficiency for each pixel within one cluster to be registered. Pixel 0 is the first hit in the cluster directly on the surface and the last pixel is the hit where the charge is deposited closest to the backside. The incident angle is \SI{78\pm2}{\degree}. Circles correspond to measurements with \SI{1000}{e} threshold, whereas squares indicate a chip threshold of \SI{800}{e}.}
\label{fig:5}
\end{figure}

Even though the overall hit efficiency of irradiated present generation pixel modules is comparatively high, local inefficiencies are clearly visible. Fig. \ref{fig:2} shows a $5\times10^{15}\;$n$_\text{eq}/$cm$^{2}$ irradiated pixel module employing the standard design biased at \SI{500}{\volt} with an overall hit efficiency of \SI{98.9\pm0.3}{\percent}. The inefficiencies are concentrated in the corners as a result of charge-sharing between four pixels and in the center right part of the pixel.
Comparing the hit efficiency map with the layout of a single pixel it becomes evident that this inefficiency is caused by the punch-through dot and the aluminum bias rail connecting the dot to the bias ring. Further investigations showed that moving the bias rail over the pixel implant instead of between the pixel columns already increases the overall hit efficiency by \SI{1}{\percent} \cite{nsavicvienna}. Despite this inefficiency not being crucial for the current pixel cell size, it will clearly be more important with smaller pixel cells. This is illustrated in Fig. \ref{fig:3} as a $50\times$\SI{50}{\square \micro \meter} cut-out of the hit efficiency map of the standard geometry pixel using the standard punch-through structure shown in Fig. \ref{fig:2}. The cut-out is composed of the left \SI{10}{\micro \meter} and the right \SI{40}{\micro \meter}.
Therefore, a further optimization of the punch-through structure is necessary. The newest layout generation, cf. Fig. \ref{fig:32}, includes a common punch-through structure where four pixels share one punch-through dot as well as optimized placement of the bias rail crossing a maximized amount of pixel implant surface. Since latest pixel prototype productions demonstrated a high yield, discarding the punch-through structure completely became conceivable. Thus, also punch-through free designs are produced and will be tested as soon as the RD53 chip becomes available.
A possibility to study the behavior of \SI{50}{\micro \meter} pitched pixels is to use the short edge of the current generation pixel cell. For this purpose highly inclined testbeam measurements are carried out. This means that the beam no longer crosses the sensor at a perpendicular incidence angle but instead nearly parallel. Fig. \ref{fig:4} illustrates the geometrical topology for a \SI{100}{\micro \meter} thick sensor. The particle enters the sensor on the surface (pixel implant site) with an angle of \SI{80}{\degree} and traverses an average of 12 pixel cells before exiting the sensor on the backside. The functional connection between incident angle and mean cluster width for different sensor thicknesses is shown in Fig. \ref{fig:42}. Highly inclined measurements mimic real detector modules from the innermost layers in the very forward and backward region. As this region is characterized by high pseudo-rapidity\footnote{$\eta \equiv -ln\left( \tan \left[ \frac{\theta}{2} \right] \right)$, $\theta$ is the angle between a track and the beam, perpendicular to the beam means $\theta=$\SI{90}{\degree}} $\eta$ values those measurements are typically labeled as high-$\eta$ studies. 
Modules of different thicknesses ranging from \SI{50}{\micro \meter} to \SI{200}{\micro \meter} were measured at a testbeam, where an incident angle of \SI{80}{\degree} was targeted. Subsequent geometrical measurements revealed a systematic effect resulting in an estimated inclination of \SI{78\pm2}{\degree}. All sensors were operated well above their full depletion voltage. Figure \ref{fig:5} shows the efficiency for each relative pixel position within the cluster. As expected, an increasing cluster width can be observed with increasing thickness. The efficiency increases with decreasing sensor thickness except for the \SI{50}{\micro \meter} thin sensor which exhibits a lower efficiency compared to the \SI{100}{\micro \meter} thick sensor. A possible explanation of this effect could be a smaller amount of charge generated in the bulk material caused by escaping $\delta$-electrons. Also, the influence of the threshold was investigated for the \SI{150}{\micro \meter} thick sensor. The efficiency increases by almost \SI{2}{\percent} when lowering the threshold from \SI{1000}{e} to \SI{800}{e}. Nevertheless, even \SI{200}{\micro \meter} thick sensors allow for an efficiency of more than \SI{90}{\percent} and track reconstruction at high-$\eta$.

\subsection{Active edge sensors}
\begin{figure}[!htp]
\centering
\includegraphics[width=8cm]{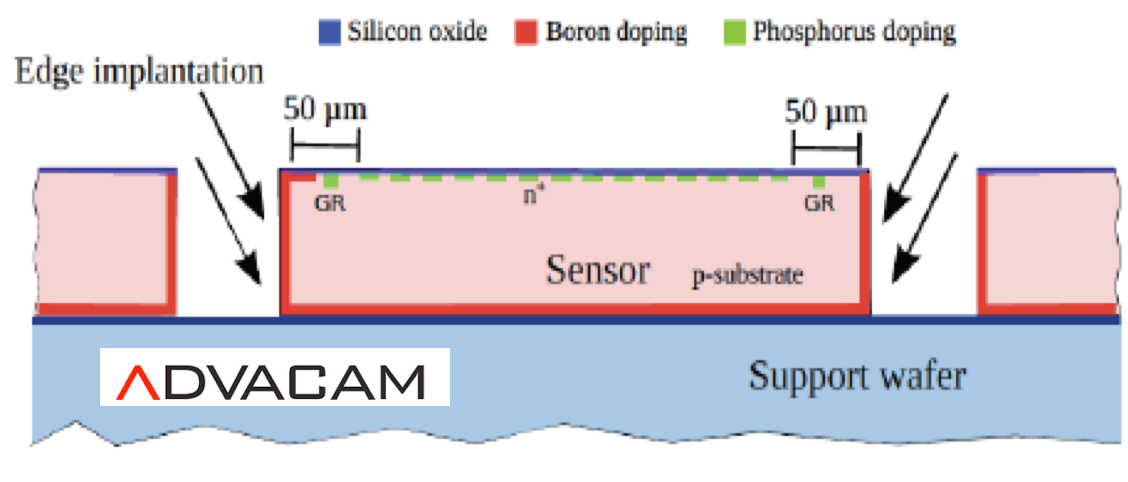}
\caption{Active edge technology used in the ADVACAM production. First the trenches are etched with the DRIE procedure on SOI wafers before the edges are implanted with an inclined boron beam. The distance between the last pixel column and the physical edge is \SI{50}{\micro \meter} for the active edge sensors and \SI{100}{\micro \meter} for the slim edge sensors.}
\label{fig:52}
\end{figure}

\begin{figure}[!htp]
\centering
\includegraphics[width=8cm]{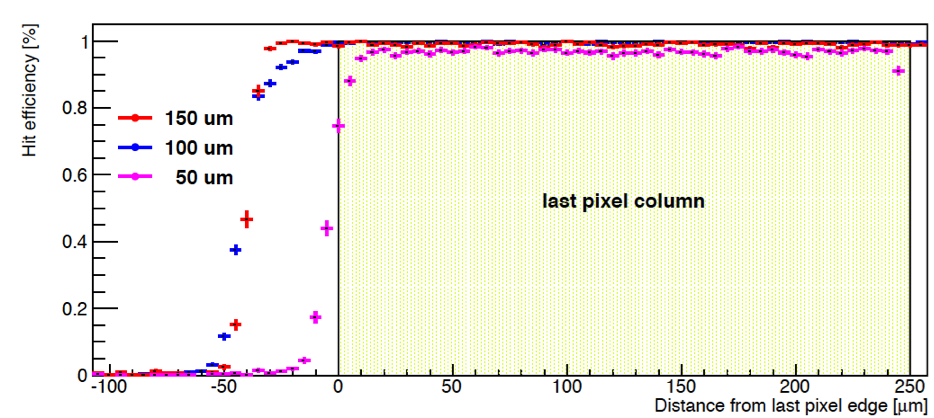}
\caption{Spatially resolved hit efficiency as a function of the relative position to the last pixel column. The measured pixels are facing the edge with their short side. A comparison is given for different thicknesses.}
\label{fig:6}
\end{figure}

\begin{figure}[!htp]
\centering
\includegraphics[width=8cm]{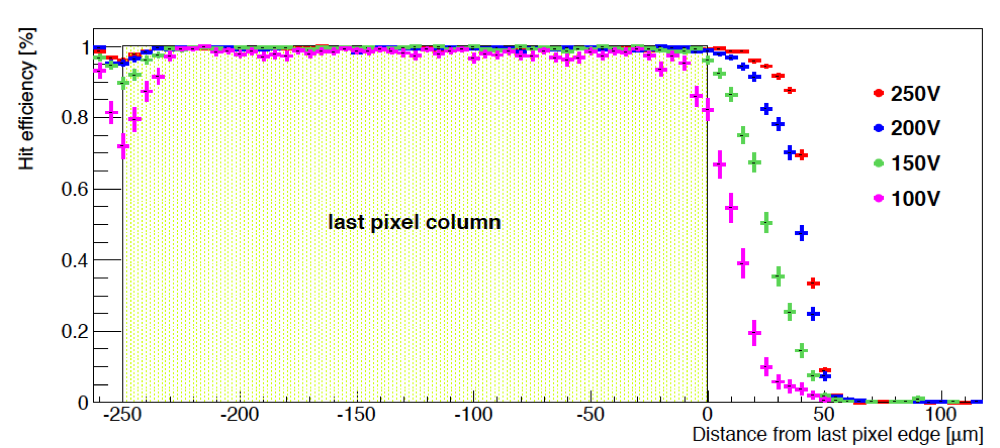}
\caption{Spatially resolved hit efficiency as a function of the position relative to the last pixel column. The measured pixels are facing the edge with their short side. A comparison is given for different voltages applied to this \SI{150}{\micro \meter} thick irradiated ($\Phi_\text{eq.}=10^{15}\,$n$_\text{eq}$/cm$^{2}$) active edge sensor.}
\label{fig:7}
\end{figure}

Pixel sensors in hybrid modules used in high energy physics are usually surrounded by a guard ring (GR) structure lowering the potential towards the edges close to ground level. An active edge process can be used as an alternative technology. Trenches created with the Deep Reactive Ion Etching (DRIE) process allow to extend the backside implantation to the sensor edges \cite{activeedge} and to create a rectifying junction reducing the defects induced by the otherwise necessary cutting. Since the electric field is extended to the side implantation, the entire edge is, in principle, fully active in the case that no bias ring is implemented.
A sketch of the technology is given in Fig. \ref{fig:52}. In the edge region of the \textit{active edge} design one floating GR is implemented in this design, resulting in a distance between the last pixel and the physical sensor edge of d$_\text{e}=$ \SI{50}{\micro \meter}. A second version of the design with d$_\text{e}=$\SI{100}{\micro \meter} a bias ring (BR) and either one or none GR is indicated in the following as slim edge sensor.
\\ \\
This production follows up a research activity initiated with VTT where FE-I3 compatible active and slim edge sensors were produced and tested \cite{weigel}. The new production is again based on silicon on insulator (SOI) wafers, but designed to fit the FE-I4 dimensions and employing sensor thicknesses of 50, 100 and \SI{150}{\micro \meter}.
Testbeam measurements were performed and the measured efficiency at the edge for different thicknesses is shown in Fig. \ref{fig:6}. The not-irradiated sensors were biased well above their full depletion voltage and the read-out chip was tuned to low thresholds of around \SI{1000}{e}. The plateau efficiency of all devices is very high (95-\SI{99}{\percent}). Nevertheless, the plateau efficiency of the \SI{50}{\micro \meter} thick sensor is smaller than that of the thicker sensors. A possible threshold effect is under investigation as it has a significant impact on the smaller charge generated in the \SI{50}{\micro \meter} thick sensor. The thresholds which will nominally be achievable with the RD53 read-out chip will be low enough to reduce threshold dependent effects for the investigated thicknesses \cite{RD53collaboration}. The \SI{100}{\micro \meter} and \SI{150}{\micro \meter} sensors are efficient up to the physical edge of the sensor.
Further measurements were carried out with an irradiated active edge sample. The first irradiation step in this case was a fluence of $10^{15}\,$n$_\text{eq}$/cm$^{2}$. The device is \SI{150}{\micro \meter} thick and the efficiency at the edge was measured for four different bias voltages ranging from 100 to \SI{250}{\volt}. The results are shown in Fig. \ref{fig:7}. The efficiency in the plateau reaches \SI{99}{\percent} and saturates already at \SI{150}{\volt}. The effect of charge sharing at \SI{-250}{\micro \meter} gets further reduced when increasing the bias voltage to \SI{250}{\volt}. Also, the efficiency at the edge increases significantly while increasing the bias voltage. At the maximal bias voltage the irradiated sensor shows a comparable performance to the not-irradiated device.


\section{Conclusion}
\label{section_results}
A novel concept of silicon pixel modules for the ATLAS ITk upgrade for HL-LHC was presented. It was shown that thin irradiated sensors with thicknesses of \SI{100}{\micro \meter} and \SI{150}{\micro \meter} perform best in terms of charge collection and hit efficiency at moderate voltages thus causing a lower power dissipation in a realistic operational scenario. During these investigations, an inefficiency caused by the punch-through structure was identified. A significant relative increase of this inefficiency is foreseen for the smaller pixel sizes of $50\times$\SI{50}{\square \micro \meter} or $25\times$\SI{100}{\square \micro \meter} of the ITk. Necessary layout changes of the punch-through structure were presented and already a \SI{1}{\percent} increase in overall hit efficiency was measured in a slightly modified design with the bias rail situated over the pixel implant.
\\ \\
High incidence angle measurements were performed with the short pixel side of FE-I4 modules to simulate the situation of $50\times$\SI{50}{\square \micro \meter} pixel cells at high pseudo-rapidities in the ITk. Measurements corresponding to $\eta=2-2.5$ were performed using sensors with thicknesses ranging from \SI{50}{\micro \meter} to \SI{200}{\micro \meter}. Sensors of all thicknesses show reasonable efficiencies allowing for track reconstruction at high pseudo-rapidities. Finally, active edge sensors were tested before and after irradiation. The feasibility of tracking up to the physical edge of the sensor for thicknesses of \SI{100}{\micro \meter} and \SI{150}{\micro \meter} has been demonstrated for not-irradiated sensors. An irradiated sensor showed that radiation induced inefficiencies can be completely compensated for by applying appropriate bias voltages.
\\ \\
In conclusion, thin planar n-in-p pixel sensors show very good results in all tested situations making them a promising candidate to cover all pixel layers of the future ITk.


\appendices

\section*{Acknowledgment}
This work has been partially performed in the framework of the CERN RD50 Collaboration. The authors thank A. Dierlamm for the irradiation at KIT, V. Cindro and I. Mandic for the irradiation at JSI. Supported by the H2020 project AIDA-2020, GA no. 654168. The authors thank the EUTelescope developer team.


\end{document}